\documentclass[a4paper,11pt]{article}
\usepackage{color,xcolor,ucs}
\usepackage[top=0.3in, bottom=0.5in, left = 0.65in, right = 0.65in]{geometry}
\usepackage[linkcolor=black,colorlinks=true,urlcolor=blue]{hyperref}
\usepackage{mathtools}

\usepackage{color,xcolor,ucs}
\usepackage{mathtools}   \usepackage{tikz} 
\usepackage{ amssymb }
\usepackage{extarrows} 
\usepackage{pgf,tikz}
\usepackage{float}
\usetikzlibrary{positioning}
\usetikzlibrary{shapes.geometric}
\usetikzlibrary{shapes.misc}
\usetikzlibrary{arrows}
\usepackage{caption}
\usepackage{mathrsfs}
\usetikzlibrary{arrows,shapes,automata,backgrounds,petri,positioning}
\usetikzlibrary{decorations.pathmorphing}
\usetikzlibrary{decorations.shapes}
\usetikzlibrary{decorations.text}
\usetikzlibrary{decorations.fractals}
\usetikzlibrary{decorations.footprints}
\usetikzlibrary{shadows}
\usetikzlibrary{calc}
\usetikzlibrary{spy}
\usepackage{amsmath}
\usepackage{array}
\usepackage{ amssymb }
\usepackage{braket}
\usepackage{qcircuit}
\usepackage{soul}
\usepackage{braket} 
\usepackage{relsize}

\usepackage{amsmath}
\usepackage{ amssymb }
\usepackage{braket}
\usepackage{qcircuit}
\usepackage{soul}
\usepackage{braket}

\title{{\LARGE Eigenvectors of Toeplitz matrices from Fisher-Hartwig symbols with greater than, or equal to, one singularity}}
\author{Pete Rigas}
\date{}

\begin{document}

\maketitle

\begin{abstract}
  Asymptotically, we analytically derive the form of eigenvectors for two Fisher-Hartwig symbols besides those which were previously investigated in a $2016$ work due to Movassagh and Kadanoff, in which the authors characterized the eigenpairs of Toeplitz matrices generated by Fisher-Hartwig symbols with one singularity. To perform such computations, we extend their methods which consists of formulating an eigenvalue problem, obtained by a Wiener-Hopf method, from which a suitable winding number is defined for passing to Fourier space, and introducing a factorization dependent upon the winding number, for other Fisher-Hartwig symbols which have previously been defined in the literature. Following the computations required for the proof after obtaining the asymptotic approximation of the eigenvectors, we provide a table from which eigenvalues of one Fisher-Hartwig symbol for different complex valued functions $b$ can be inferred. \footnote{\textit{{Keywords}}: Random matrix theory, Toepltiz matrices, Fisher-Hartwig symbol, complex plane, Wiener-Hopf} \footnote{\textbf{MSC Class}: 82D02; 81V02}
\end{abstract}

\section{Introduction}

\subsection{Overview}

Fisher-Hartwig theory has long been known for connections with Toeplitz matrices, as previous works have identified approaches for computing determinants {\color{blue}[3,4]}, quantifying matrix norms 
{\color{blue}[5]}, analysis of piecewise continuity {\color{blue}[6]}, and singularities {\color{blue}[1,8]}, of Fisher-Hartwig symbols, entangelment entropy {\color{blue}[10]}, and spectral theory {\color{blue}[11]}, amongst others {\color{blue}[15,16,17]}. To further investigate characteristics of the Fisher-Hartwig theory, we answer one open question raised in {\color{blue}[14]} pertaining to the computation of eigenvalues, eigenfunctions, and momenta of eigenvalues of Toeplitz matrices generated by Fisher-Hartwig symbols. In comparison to {\color{blue}[14]} which provided analytical expressions for the eigenvalues and eigenfunctions of Toeplitz matrices from Fisher-Hartwig symbols of a type previously considered in {\color{blue}[7,9]}, we apply the Wiener-Hopf method to Fisher-Hartwig symbols of two other types which have been previously defined in the literature. After having identified Fisher-Hartwig symbols for which the Wiener-Hopf method has not been applied, we obtain the analytical form of eigenvectors for other classes of Toeplitz matrices rather than those previously analyzed in {\color{blue}[14]}. As expected, the Wiener-Hopf approach due to {\color{blue}[14]} from which the eigenpairs of Toeplitz matrices can be analyzed is reliant upon the analytical form of the symbol. In order to apply this method in new situations, we make use of symbols which are defined in {\color{blue}[4,7]}.

\subsection{Paper organization}

We provide a description of previous results obtained for Fisher-Hartwig symbols in {\color{blue}[14]}, followed by  conditions placed on the definition of the symbol. Following the statement of these properties, we identify two new Fisher-Hartwig symbols from the literature which can be analyzed with the Wiener-Hopf method, which we begin implementing in \textit{2}. From symbols of each type which are introduced in the next section, \textit{1.3}, we address Question $1$ raised by the authors in the final section of {\color{blue}[14]}, in which the authors suspect that their method for asymptotically determining the eigenvectors of a Toeplitz matrix, generated by singular Fisher-Hartwig symbols, can be applied to other symbols that have not been analyzed previously in this light. 

\subsection{Analysis of the spectrum of Toeplitz matrices from the Fisher-Hartwig symbol}

For the unit circle,

\begin{align*}
  \textbf{s} \equiv \big\{ z \in \textbf{C} : |z| = 1 \big\}      \text{ } \text{ , } 
\end{align*}

\noindent the Fisher-Hartwig symbol, with generating function, for some $n>0$,

\begin{align*}
 a \big( z \big) = \overset{n-1}{\underset{k = - ( n-1)}{\sum}}  t_k z^k \text{ } \text{ , } 
\end{align*}

\noindent is of the form,

\begin{align*}
  a \big( z \big) = e^{V(z)} z^{\sum_{j=0}^m \beta_j } {\prod}_{j=0}^m \big| z - \bar{z}_j \big|^{2 \alpha_j} g_{z_j , \beta_j} (z ) \bar{z}_j^{-\beta_j}   \text{ } \text{ , } 
\end{align*}

\noindent for smooth, complex valued, $V(z)$, with,

\begin{align*}
  z = e^{-i p} \text{ } \text{ , } 0 \leq p \leq 2 \pi  \text{ } \text{ , } 
\end{align*}

\noindent subject to the conditions,

\[
g_{z_j , \beta_j} (z )  \equiv  \text{ } 
\left\{\!\begin{array}{ll@{}>{{}}l} e^{i \pi \beta_j } & \text{ } \text{ , }  0 \leq \mathrm{arg}\big( z\big) < p_j  \text{ } \\
   e^{ - i \pi \beta_j } &    \text{ } \text{ , } p_j \leq \mathrm{arg} \big( z \big) < 2 \pi      \text{ }  \\
\end{array}\right.
\]

\noindent on $g_{z_j , \beta_j} (z )$, in addition to the conditions, 

\begin{align*}
   \mathrm{Re} \big( \alpha_j \big) > - \frac{1}{2}      \text{ } \text{ , }  \beta_j \in \textbf{C} \text{ } \text{ , } 
\end{align*}

\noindent on the parameters $\alpha_j$ and $\beta_j$, for $j = 0 , \cdots , m$. Also, denote $v \big( \lambda , a \big)$ as the winding number of $a \big( \textbf{s} \big)$ around $\lambda \in \textbf{C}$. From the symbol, the $j,k$th entry of the Toeplitz matrix can be defined with,

\begin{align*}
     t_{j,k} \equiv t_{j-k} \equiv \frac{1}{2 \pi i} \int_{\textbf{s}}                 \frac{a \big( z \big)}{z^{j-k+1}}     \text{ }     \mathrm{d} z \text{ }       \equiv \big( -1 \big)^{j-k} \frac{\Gamma \big( 2 \alpha + 1 \big) }{\Gamma \big( \alpha + \beta + 1 - j + k  \big) \Gamma \big( \alpha - \beta + 1 + j - k  \big)}                           \text{ } \text{ , } 
\end{align*}

\noindent over the boundary contour $\textbf{s}$ in the complex plane. Equipped with each $t_{j,k}$, the \textit{Toeplitz matrix} is a matrix whose entries are constant along the diagonals, and whose off-diagonal entries decrease with respect to the distance from the diagonal. Besides this description in words, one such $n \times n$ matrix that is \textit{Toeplitz} is given by $T_n \big( a \big) \equiv \{ t_{j-k} \}_{0 \leq j , k \leq n-1}$, in which the entries along the diagonal are given by $t_0$, while entries below the diagonal are given by $t_1 , \cdots , t_{(n-1)}$ corresponding to each diagonal \textit{below} the main diagonal. Conversely, above the main diagonal, the entries along each diagonal are given by $t_{-1} , \cdots , t_{-(n-1)}$.

Asymptotically, the momenta, eigenvalues, and eigenvectors, of the $l$th item in the spectrum of a \textit{Toeplitz matrix} are given by,

\begin{align*}
     p^l   \equiv \frac{2 \pi l}{n} + i \big( 2 \alpha + 1 \big) \frac{\mathrm{log} n }{n} + \mathrm{O} \big( n^{-1} \big)     \text{ } \text{ , } \\   E^l \equiv a \big[ \mathrm{exp} \big(  - i p^l    \big)  \big]   + \mathrm{o} \big( n^{-1} \big)      \text{ } \text{ , } \\           \psi^l_j \propto \mathrm{exp} \big( i p^l j \big)    \text{ } \text{ , } 
\end{align*}

\noindent respectively {\color{blue}[13]}. Previously, in {\color{blue}[14]}, under the assumption that the symbol is singular, in which the symbol takes the form,

\begin{align*}
    a \big( z \big) =  \big( - 1 \big)^{\alpha + \beta }  \big( \frac{z-1}{z} \big)^{2 \alpha} z^{\beta}  \equiv  \big( z - 1 \big)^{2 \alpha} z^{\beta - \alpha} \mathrm{exp} \big[   - i \big( \alpha + \beta \big) \pi   \big]   \text{ } \text{ , } 
\end{align*}

\noindent the authors approximate the eigenvalues with,

\begin{align*}
   E^l \approx \big( - 1 \big)^{\beta} 4^{\alpha} \mathrm{sin}^{2 \alpha} \big( \frac{\pi l}{n} \big) \mathrm{exp} \big( - \frac{2 i \pi \beta l}{n}  \big)  \text{ } \text{ . } 
\end{align*}

\noindent For Fisher-Hartwig symbols besides the singular symbols given by $a \big( z \big)$ above, it is also possible that such a quantity takes the form,

\begin{align*}
     a^{(\mathrm{Type\text{ }I})} \big( z \big)  = \big( 2 - z - \frac{1}{z} \big)^{\frac{\alpha}{2}} \big( - z \big)^{\beta}            \text{ } \text{ , }  \tag{$\mathrm{Type\text{ } I}$}
\end{align*}

\noindent as presented in {\color{blue}[7]}, with,

\begin{align*}
    0 < \frac{\alpha}{2} < - \beta < 1   \text{ } \text{ , } 
\end{align*}

\noindent corresponding to other symbols besides the singular ones, with eigenvalues approximately given by,

\begin{align*}
E^l =   \bigg[ 2 - \mathrm{exp}  \big( - i p^l    \big)  - \mathrm{exp}  \big( - i p^l    \big)^{-1} \bigg]^{\frac{\alpha}{2}} \big( -  \mathrm{exp}  \big( - i p^l    \big) 
 \big)^{\beta}   + \mathrm{o} \big(n^{-1} \big)  \approx           \bigg[ 2 - \mathrm{exp}  \big( - i p^l    \big)  - \mathrm{exp}  \big( - i p^l    \big)^{-1} \bigg]^{\frac{\alpha}{2}} \times \cdots \\   \big( -1  
 \big)^{\beta}  \mathrm{exp}^{\beta}  \big( - i p^l  \big)  \text{ } \text{ , }  \end{align*}
 
\noindent from which additional rearrangements yield, approximately,
 
 \begin{align*}
 \bigg[ 2 - \mathrm{exp}  \bigg( - i \big[  \frac{2 \pi l}{n}  + \mathrm{O} \big( n^{-1} \big) \big]   \bigg)  - \mathrm{exp}  \bigg( - i \big[ \frac{2 \pi l}{n}  + \mathrm{O} \big( n^{-1} \big) \big]     \bigg)^{-1} \bigg]^{\frac{\alpha}{2}}   \big( - 1 \big)^{\beta} \mathrm{exp}^{\beta}   \bigg( - i \big[  \frac{2 \pi l}{n}  + \mathrm{O} \big( n^{-1} \big)\big]     \bigg) 
         \text{ } \text{ , } 
\end{align*}

\noindent upon ignoring the $\frac{\mathrm{log n}}{n}$ term in $p^l$, in which,

\begin{align*}
   p^l \approx \frac{2 \pi l}{n}  + \mathrm{O} \big(  n^{-1} \big)     \text{ } \text{ , }
\end{align*}

\noindent hence implying that the eigenvalues are approximately,

\begin{align*}
 E^{l,\mathrm{(Type\text{ }I)}} \approx      \bigg[ 2 - \mathrm{exp}  \big(  \frac{-2 i \pi l}{n}   \big)    - \mathrm{exp}   \big( \frac{-2 i \pi l}{n}   \big)^{-1} \bigg]^{\frac{\alpha}{2}}   \big( -1  
 \big)^{\beta}  \mathrm{exp}^{\beta}  \big( \frac{-2 i \pi l}{n}  \big)          \text{ } \text{ . } 
\end{align*}

\noindent Alternatively, the symbol could assume the form, for $|z|=1$,

\begin{align*}
a^{(\mathrm{Type\text{ }II})} \big( z \big)  = \big( 1 - \frac{z_0}{z} \big)^{\delta} \big( 1 - \frac{z}{z_0}     \big)^{\gamma} b \big( z \big)   \overset{\text{countably}}{\overset{\text{many jump}}{\overset{\text{discontinuities\text{ }} z_j}{\Longleftrightarrow}}}    u_{\alpha_j} \big( z - z_j \big)       \overset{j}{\underset{i=1}{\prod}}     t_{\beta_j} \big(   z - z_j    \big) \text{ }      b \big( z \big)    \text{ } \text{ , }       \tag{$\mathrm{Type \text{ } II}$} 
\end{align*}

\noindent for some fixed $z_0$ on $\textbf{s}$, complex parameters $\delta$ and $\gamma$ subject to the condition,

\begin{align*}
 \mathrm{Re} \big( \delta \big) + \mathrm{Re} \big( \gamma \big) > - 1   \text{ } \text{ , } 
\end{align*}

\noindent and for $b$ sufficiently smooth which does not vanish on $\textbf{s}$, corresponding to symbols presented in {\color{blue}[4]}, with eigenvalues,

\begin{align*}
    E^l =  \bigg[ 1 - \frac{z_0}{\mathrm{exp} \big( - i p^l \big) } \bigg]^{\delta} \bigg[ 1 - \frac{\mathrm{exp} \big( - i p^l \big)}{z_0}     \bigg]^{\gamma} b \bigg( \mathrm{exp} \big( - i p^l \big) \bigg)     + \mathrm{o} \big( n^{-1} \big)  \approx   \bigg[ 1 - \frac{z_0}{\mathrm{exp} \big( - i p^l \big) } \bigg]^{\delta} \bigg[ 1 - \frac{\mathrm{exp} \big( - i p^l \big)}{z_0}     \bigg]^{\gamma} \times \cdots \\  b \bigg( \mathrm{exp} \big( - i p^l \big) \bigg)      \text{ , } 
\end{align*}

\noindent from which making the same truncation on $p^l$ as in the previous computation yields, approximately,

\begin{align*}
  E^{l,\mathrm{(Type\text{ }II)}}  \approx      \bigg[ 1 - \frac{z_0}{\mathrm{exp} \big( \frac{- 2 i \pi l}{n} \big) } \bigg]^{\delta} \bigg[ 1 - \frac{\mathrm{exp} \big( \frac{- 2 i \pi l}{n} \big)}{z_0}     \bigg]^{\gamma}  b \bigg( \mathrm{exp} \big( \frac{- 2 i \pi l}{n} \big) \bigg)           \text{ } \text{ . } 
\end{align*}

\noindent A final symbol of interest takes the form, {\color{blue}[16]},

\begin{align*}
      a^{(\mathrm{Type\text{ }III})} \big( z \big)  =  \overset{M}{\underset{j=1}{\prod}} \big| z_j \big|^{2 \alpha_j }    \bigg| \text{ } \frac{z}{z_j} - 1 \text{ } \bigg|^{2 \alpha_j} \text{ }  c \big( z \big)        \text{ } \text{ , } 
\end{align*}

\noindent with $-\frac{1}{2} < \alpha_j < \frac{1}{2}$, and $c$ sufficiently smooth. The corresponding eigenvalues take the form,

\begin{align*}
    E^{l,\mathrm{(Type\text{ }III)}} =         a^{(\mathrm{Type\text{ }III})} \bigg( \mathrm{exp} \big( - i p^l \big) \bigg)  + \mathrm{o} \big( n^{-1} \big)           =   \overset{M}{\underset{j=1}{\prod}}   \bigg(  \big| \mathrm{exp} \big( - i p^l_j \big) \big|^{2 \alpha_j }    \bigg| \text{ }   \frac{\mathrm{exp} \big( - i p^l \big) }{\mathrm{exp} \big( - i p^l_j \big) }  - 1 \text{ } \bigg|^{2 \alpha_j} \bigg)\times \cdots \\  \text{ }  c \bigg( \mathrm{exp} \big( - i p^l \big)  \bigg)        + \mathrm{o} \big( n^{-1} \big)      \text{ } \text{ , } 
\end{align*}

\noindent which approximately yields,

\begin{align*}
    E^{l,\mathrm{(Type\text{ }III)}}  \approx  \overset{M}{\underset{j=1}{\prod}} \bigg( \big| \mathrm{exp} \big( - i p^l_j \big) \big|^{2 \alpha_j }   \bigg| \text{ }   \frac{\mathrm{exp} \big( - i p^l \big) }{\mathrm{exp} \big( - i p^l_j \big) }  - 1 \text{ } \bigg|^{2 \alpha_j} \bigg)  \text{ }  c \bigg( \mathrm{exp} \big(  \frac{- 2 i \pi l}{n} \big)  \bigg)            \text{ } \text{ , } 
\end{align*}

\noindent From the symbols of each type defined in this section, the eigenvectors asymptotically satisfy the following relations.

\bigskip

\noindent \textbf{Theorem} (\textit{applying results from} {\color{blue}[15]} \textit{to obtain eigenpairs for other Fisher-Hartwig symbols}). For Fisher-Hartwig symbols of any of the three types, asymptotically the eigenvectors satisfy,

\begin{align*}
  \psi^{l, (\mathrm{Type\text{ } I})}_j - B \big( j + 1 \big)^{-2\alpha -1} \sim A n^{-2 \alpha -1} z_{\mathrm{crit}}^{-1, (\mathrm{Type\text{ } I})}    \text{ } \text{ , } 
\end{align*}

\noindent for $E^{l,(\mathrm{Type\text{ } I})}$,

\begin{align*}
  \psi^{l, (\mathrm{Type\text{ } II})}_j - B \big( j + 1 \big)^{-2\alpha -1} \sim A n^{-2 \alpha -1} z_{\mathrm{crit}}^{-1, (\mathrm{Type\text{ } II})}    \text{ } \text{ , } 
\end{align*}

\noindent for $E^{l,(\mathrm{Type\text{ } II})}$, and, 

\begin{align*}
  \psi^{l, (\mathrm{Type\text{ } III})}_j - B \big( j + 1 \big)^{-2\alpha -1} \sim A n^{-2 \alpha -1} z_{\mathrm{crit}}^{-1, (\mathrm{Type\text{ } III})}    \text{ } \text{ , } 
\end{align*}

\noindent for $E^{l,(\mathrm{Type\text{ } III})}$, for constants $A$ and $B$ depending on $\alpha$, $\beta$, and each $E^l$.

\bigskip

\noindent With the three symbols, two of which have previously been analyzed, and approximate expressions for $E^l$, in the next section we execute the Wiener-Hopf method and following steps.

\section{The Wiener-Hopf Method for Symbols of Types I, II $\&$ III}

\noindent The current section containing the proof is a three-pronged strategy, involving stating the eigenvalue problem, passing to Fourier space, and asymptotically evaluating a contour integral which is a function of the winding number.

\subsection{Eigenvalue problem formulation}

\noindent \textit{Proof}. To apply the Wiener-Hopf method, one must formulate an eigenvalue problem, of the form,

\begin{align*}
     \overset{n}{\underset{j=0}{\sum}}  W_{i-j} \psi^l_j = 0 \text{ } \text{ , } \tag{$\mathrm{Equation\text{ }1}$}
\end{align*}

\noindent for $i\geq 0$, corresponding to the right eigenvector, and,

\begin{align*}
        \overset{n}{\underset{j=0}{\sum}} \widetilde{\psi^l_j} W_{j-i} = 0   \text{ } \text{ , } 
\end{align*}

\noindent corresponding to the left eigenvector, also for $i \geq 0$. The quantity $W$ appearing in each summation above is, for a Toeplitz matrix $T$,

\begin{align*}
     W = T - E^l \textbf{I}   \text{ } \text{ , } 
\end{align*}

\noindent for the identity matrix of size $n$, $\textbf{I}$, implying that $W$ assumes the form,

\begin{align*}
  W^{(\mathrm{Type \text{ } I})} \approx      T -   \bigg[   \bigg[ 2 - \mathrm{exp}  \big(  \frac{-2 i \pi l}{n}   \big)    - \mathrm{exp}   \big( \frac{-2 i \pi l}{n}   \big)^{-1} \bigg]^{\frac{\alpha}{2}}   \big( -1  
 \big)^{\beta}  \mathrm{exp}^{\beta}    \big(  \frac{- 2 i \pi l}{n}  \big)   \bigg]   \textbf{I}      \text{ } \text{ , } 
\end{align*}

\noindent for Fisher-Hartwig symbols of $\mathrm{Type\text{ } I}$, and,

\begin{align*}
   W^{(\mathrm{Type \text{ } II})}  \approx    T -        \bigg[ 1 - \frac{z_0}{\mathrm{exp} \big( \frac{- 2 i \pi l}{n} \big) } \bigg]^{\delta} \bigg[ 1 - \frac{\mathrm{exp} \big( \frac{- 2 i \pi l}{n} \big)}{z_0}     \bigg]^{\gamma}  b \bigg( \mathrm{exp} \big( \frac{- 2 i \pi l}{n} \big) \bigg)          \textbf{I}   \text{ } \text{ , } 
\end{align*}

\noindent for Fisher-Hartwig symbols of $\mathrm{Type\text{ } II}$. Finally, $W$ for the remaining type of Fisher-Hartwig symbols takes the form,

\begin{align*}
 W^{(\mathrm{Type \text{ } III})}  \approx T -    \overset{M}{\underset{j=1}{\prod}}   \bigg(  \big| \mathrm{exp} \big( - i p^l_j \big) \big|^{2 \alpha_j }    \bigg| \text{ }   \frac{\mathrm{exp} \big( - i p^l \big) }{\mathrm{exp} \big( - i p^l_j \big) }  - 1 \text{ } \bigg|^{2 \alpha_j} \bigg) \text{ }  c \bigg( \mathrm{exp} \big( - i p^l \big)  \bigg)      \textbf{I}  \text{ } \text{ , } 
\end{align*}

\noindent With $W$ for $(\mathrm{Type\text{ } I})$ and $(\mathrm{Type\text{ } II})$ symbols introduced in \textit{1.3}, one must also stipulate,

\begin{align*}
 \overset{+\infty}{\underset{j=0
}{\sum}} \big| \psi^{l , \mathrm{(Type\text{ } I)}} \big| < + \infty  \text{ } \text{ , } \end{align*} 

\noindent for the singly infinite series, and,

\begin{align*}\overset{+\infty}{\underset{j = -\infty}{\sum}}  \big| W^{\mathrm{(Type\text{ } I)}}_j  \big| = \overset{+\infty}{\underset{j = -\infty}{\sum}}  \big| T_j - E^{l,\mathrm{(Type\text{ }I)}}_j \textbf{I}  \big| \approx \overset{+\infty}{\underset{j = -\infty}{\sum}}  \big|  T_j -      \bigg[ 1 - \frac{z_0}{\mathrm{exp} \big( \frac{- 2 i \pi j}{n} \big) } \bigg]^{\delta} \bigg[ 1 - \frac{\mathrm{exp} \big( \frac{- 2 i \pi j}{n} \big)}{z_0}     \bigg]^{\gamma}  b \big( \mathrm{exp} \big( \frac{- 2 i \pi j}{n} \big) \big)          \textbf{I}       \big| \text{ } \text{ , } 
\end{align*}

\noindent for the doubly infinite series, where $\psi^{l , \mathrm{Type\text{ } I}}$ denotes the $l$ th left eigenvector of $(\mathrm{Type\text{ } I})$, and,

\begin{align*}
 \overset{+\infty}{\underset{j=0
}{\sum}} \big| \psi^{l , \mathrm{(Type\text{ } II)}} \big| < + \infty  \text{ } \text{ , }  \end{align*}

\noindent for the singly infinite series, 

\begin{align*}\overset{+\infty}{\underset{j = -\infty}{\sum}}  \big| W^{\mathrm{(Type\text{ } II)}}_j  \big|  =  \overset{+\infty}{\underset{j = -\infty}{\sum}}  \big| T - E^{l,\mathrm{(Type\text{ }II)}}_j \textbf{I}\big| \approx \overset{+\infty}{\underset{j = -\infty}{\sum}}  \big|  T_j -   \big[ 1 - \frac{z_0}{\mathrm{exp} \big( \frac{- 2 i \pi j}{n} \big) } \big]^{\delta} \big[ 1 - \frac{\mathrm{exp} \big( \frac{- 2 i \pi j}{n} \big)}{z_0}     \big]^{\gamma}  b \big( \mathrm{exp} \big( \frac{- 2 i \pi j}{n} \big) \big)     \textbf{I}      \big|   \text{ } \text{ , } 
\end{align*}

\noindent for the doubly infinite series, where $\psi^{l , (\mathrm{Type\text{ } II})}$ denotes the $l$ th left eigenvector of $(\mathrm{Type\text{ } II})$. For $(\mathrm{Type \text{ } III})$, the singly infinite series takes the form,

\begin{align*}
    \overset{+\infty}{\underset{j=0
}{\sum}} \big| \psi^{l , \mathrm{(Type\text{ } III)}} \big| < + \infty \text{ } \text{ , } 
\end{align*}

\noindent while the doubly infinite series takes the form,

\begin{align*}
  \overset{+\infty}{\underset{j = -\infty}{\sum}}  \big| W^{\mathrm{(Type\text{ } III)}}_j  \big|  =  \overset{+\infty}{\underset{j = -\infty}{\sum}}  \big| T - E^{l,\mathrm{(Type\text{ }III)}}_j \textbf{I}\big| \approx \overset{+\infty}{\underset{j = -\infty}{\sum}} \bigg| T -  \bigg[  \text{ } \overset{M}{\underset{j=1}{\prod}} \bigg(  \big| \mathrm{exp} \big( - i p^l_j \big) \big|^{2 \alpha_j }    \bigg| \text{ }   \frac{\mathrm{exp} \big( - i p^l \big) }{\mathrm{exp} \big( - i p^l_j \big) }  - 1 \text{ } \bigg|^{2 \alpha_j} \bigg) \text{ } \times \cdots \\  c \bigg( \mathrm{exp} \big( - i p^l \big)  \bigg)    \bigg]   \textbf{I} \bigg| \text{ } \text{ . } 
\end{align*}

\subsection{Passing to Fourier space}

\noindent The first equation in \textit{2.1} can be rewritten as,

\begin{align*}
      \overset{+ \infty}{\underset{j = -\infty}{\sum}} W^{(\mathrm{Type\text{ } I)}}_{i-j} \psi^{l , (\mathrm{Type\text{ } I})}_j =      \Theta \big( - i \big)   \overset{+\infty}{\underset{j=0}{\sum}}        \widetilde{\psi^{l , \mathrm{(Type\text{ } I})}_j}     W^{(\mathrm{Type\text{ } I)}}_{j-i}       \text{ } \text{ , } 
\end{align*}

\noindent for $(\mathrm{Type\text{ } I})$,

\begin{align*}
      \overset{+ \infty}{\underset{j = -\infty}{\sum}} W^{(\mathrm{Type\text{ } II)}}_{i-j} \psi^{l , (\mathrm{Type\text{ } II)}}_j =  \Theta \big( - i \big)   \overset{+\infty}{\underset{j=0}{\sum}}        \widetilde{\psi^{l , \mathrm{(Type\text{ } II})}_j}     W^{(\mathrm{Type\text{ } II)}}_{j-i}         \text{ } \text{ , } 
\end{align*}

\noindent for $(\mathrm{Type\text{ } II})$, and,

\begin{align*}
     \overset{+ \infty}{\underset{j = -\infty}{\sum}} W^{(\mathrm{Type\text{ } III)}}_{i-j} \psi^{l , (\mathrm{Type\text{ } III)}}_j =  \Theta \big( - i \big)   \overset{+\infty}{\underset{j=0}{\sum}}        \widetilde{\psi^{l , \mathrm{(Type\text{ } III})}_j}     W^{(\mathrm{Type\text{ } III)}}_{j-i}        \text{ } \text{ , } 
\end{align*}

\noindent for $(\mathrm{Type\text{ } III})$, for $i \in \textbf{Z}$, and for the \textit{heaviside function},

\[ \Theta \big( i \big)\equiv \left\{\!\begin{array}{ll@{}>{{}}l}                0 \text{ if } i < 0 \text{ } \text{ , }  \\        1 \text{ if } i \geq 0     
  \text{ } \text{ . } \\
\end{array}\right.
\]

\noindent Taking the Fourier transform of each equality above yields,

\begin{align*}
    \mathcal{W}^{(\mathrm{Type\text{ } I})} \big( z \big) \widetilde{\Psi^{l,(\mathrm{Type\text{ } I})}}_j  =    \widetilde{\Psi^{l,(\mathrm{Type\text{ } I})}}_i     \text{ } \text{ , } 
\end{align*}

\noindent corresponding to $(\mathrm{Type\text{ }I})$, 

\begin{align*}
    \mathcal{W}^{(\mathrm{Type\text{ } II})} \big( z \big)  \widetilde{\Psi^{l,(\mathrm{Type\text{ } II})}}_j  =    \widetilde{\Psi^{l,(\mathrm{Type\text{ } II})}}_i    \text{ } \text{ , } 
\end{align*}

\noindent corresponding to $(\mathrm{Type\text{ }II})$, and,

\begin{align*}
    \mathcal{W}^{(\mathrm{Type\text{ } III})} \big( z \big)  \widetilde{\Psi^{l,(\mathrm{Type\text{ } III})}}_j  =    \widetilde{\Psi^{l,(\mathrm{Type\text{ } III})}}_i   \text{ } \text{ , }
\end{align*}

\noindent corresponding to $(\mathrm{Type\text{ }III})$. In Fourier Space, $\widetilde{\Psi^{l,(\mathrm{Type\text{ } I})}}_j$ and $\widetilde{\Psi^{l,(\mathrm{Type\text{ } I})}}_i$ are related in that $\widetilde{\Psi^{l,(\mathrm{Type\text{ } I})}}_j$ is nonzero for $i \geq 0$, and vanishes for $i<0$, while $\widetilde{\Psi^{l,(\mathrm{Type\text{ } I})}}_i$ vanishes for $i \geq 0$ and is nonzero for $i < 0$. The same observation applies for $\widetilde{\Psi^{l,(\mathrm{Type\text{ } II})}}_j$ and $\widetilde{\Psi^{l,(\mathrm{Type\text{ } II})}}_i$, and also for $\widetilde{\Psi^{l,(\mathrm{Type\text{ } III})}}_j$ and $\widetilde{\Psi^{l,(\mathrm{Type\text{ } III})}}_i$. From each equation, the winding number $\mathcal{W} \big( z \big)$ of each type is respectively given by,

\begin{align*}
   \widetilde{ W^{(\mathrm{Type \text{ } I})} }  \equiv  \mathcal{W}^{(\mathrm{Type\text{ } I})} \big( z \big) \approx   a^{(\mathrm{Type\text{ }I})} \big( z \big)    - \bigg[ 2 - \mathrm{exp}  \big(  \frac{-2 i \pi l}{n}   \big)    - \mathrm{exp}   \big( \frac{-2 i \pi l}{n}   \big)^{-1} \bigg]^{\frac{\alpha}{2}}   \big( - 1 \big)^{\beta} \mathrm{exp}^{\beta}   \big(  \frac{- 2 i \pi l}{n}     
 \big)   \text{ } \text{ , } \\
    \widetilde{ W^{(\mathrm{Type \text{ } II})} }  \equiv    \mathcal{W}^{(\mathrm{Type\text{ } II})} \big( z \big) \approx  a^{(\mathrm{Type\text{ }II})} \big( z \big)   -     \bigg[ 1 - \frac{z_0}{\mathrm{exp} \big( \frac{- 2 i \pi l}{n} \big) } \bigg]^{\delta} \bigg[ 1 - \frac{\mathrm{exp} \big( \frac{- 2 i \pi l}{n} \big)}{z_0}     \bigg]^{\gamma}  b \bigg( \mathrm{exp} \big( \frac{- 2 i \pi l}{n} \big) \bigg)        \text{ } \text{ , } \\     \widetilde{ W^{(\mathrm{Type \text{ } III})} }  \equiv    \mathcal{W}^{(\mathrm{Type\text{ } III})} \big( z \big) \approx  a^{(\mathrm{Type\text{ }III})} \big( z \big)   -               \overset{M}{\underset{j=1}{\prod}}  \bigg(  \big| \mathrm{exp} \big( - i p^l_j \big) \big|^{2 \alpha_j }    \bigg| \text{ }   \frac{\mathrm{exp} \big( - i p^l \big) }{\mathrm{exp} \big( - i p^l_j \big) }  - 1 \text{ } \bigg|^{2 \alpha_j} \bigg) \text{ }  c \bigg( \mathrm{exp} \big( \frac{-2 i \pi l}{n} \big)  \bigg)       \text{ } \text{ . } 
\end{align*}

\noindent With the Fourier space representation $\widetilde{\Psi^l}$ of $\psi_l$, a Taylor series approximation of $\widetilde{\Psi^{l,(\mathrm{Type\text{ } I})}}_j$, and of $\widetilde{\Psi^{l,(\mathrm{Type\text{ } II})}}_j$, are of the form,

\begin{align*}
    \widetilde{\Psi^{l,(\mathrm{Type\text{ } I})}}_j    =    \underset{n \geq 0}{\sum}   a^{(\mathrm{Type\text{ } I})}_n z^n        \text{ } \text{ , }\\\widetilde{\Psi^{l,(\mathrm{Type\text{ } II})}}_j =     \underset{n \geq 0}{\sum}   a^{(\mathrm{Type\text{ } II})}_n   z^n     \text{ } \text{ , } \\ \widetilde{\Psi^{l,(\mathrm{Type\text{ } III})}}_j =     \underset{n \geq 0}{\sum}   a^{(\mathrm{Type\text{ } III})}_n   z^n    \text{ } \text{ , } 
\end{align*}

\noindent with $\underset{n \geq 0}{\sum} | a^{(\mathrm{Type\text{ } I})}_n | , \underset{n \geq 0}{\sum} | a^{(\mathrm{Type\text{ } II})}_n | ,\underset{n \geq 0}{\sum} | a^{(\mathrm{Type\text{ } III})}_n | < + \infty$. With the winding number $\mathcal{W}$, one can also introduce the quantities,

\begin{align*}
     v^{(\mathrm{Type\text{ } I})}     \equiv       \big( 2 \pi i \big)^{-1} \mathrm{log} \big[   \frac{\mathcal{W}^{(\mathrm{Type\text{ } I})} \big( \mathrm{exp} \big( 2 \pi i \big)  \big) }{ \mathcal{W}^{(\mathrm{Type\text{ } I})} \big( 1 \big)  }    \big]      =   \big( 2 \pi i \big)^{-1} \mathrm{log} \big[   \frac{ a^{(\mathrm{Type\text{ } I})}\big( \mathrm{exp} \big( 2 \pi i \big)  \big)  -   E^{l,\mathrm{(Type\text{ }I)}}   -\mathrm{o} \big( n^{-1} \big)  }{ a^{(\mathrm{Type\text{ } I})}\big( \mathrm{exp} \big(1  \big)  \big) -   E^{l,\mathrm{(Type\text{ }I)}}   -\mathrm{o} \big( n^{-1} \big)  }    \big]  \text{ } \text{ , } \\           v^{(\mathrm{Type\text{ } II})}   \equiv        \big( 2 \pi i \big)^{-1}   \mathrm{log} \big[     \frac{\mathcal{W}^{(\mathrm{Type\text{ } II})} \big( \mathrm{exp} \big( 2 \pi i \big)  \big)  }{ \mathcal{W}^{(\mathrm{Type\text{ } II})} \big( 1 \big) }       \big]   =  \big( 2 \pi i \big)^{-1} \mathrm{log} \big[   \frac{ a^{(\mathrm{Type\text{ } II})}\big( \mathrm{exp} \big( 2 \pi i \big)  \big)  -   E^{l,\mathrm{(Type\text{ }II)}}   -\mathrm{o} \big(n^{-1}\big)  }{ a^{(\mathrm{Type\text{ } II})}\big( \mathrm{exp} \big(1  \big)  \big) -   E^{l,\mathrm{(Type\text{ }II)}}   -\mathrm{o} \big( n^{-1} \big)  }    \big] \text{ } \text{ , } \\    v^{(\mathrm{Type\text{ } III})}   \equiv   \big( 2 \pi i \big)^{-1}   \mathrm{log} \big[     \frac{\mathcal{W}^{(\mathrm{Type\text{ } III})} \big( \mathrm{exp} \big( 2 \pi i \big)  \big)  }{ \mathcal{W}^{(\mathrm{Type\text{ } III})} \big( 1 \big)  }       \big]        =   \big( 2 \pi i \big)^{-1} \mathrm{log} \big[   \frac{ a^{(\mathrm{Type\text{ } III})}\big( \mathrm{exp} \big( 2 \pi i \big)  \big)  -   E^{l,\mathrm{(Type\text{ }III)}}   -\mathrm{o} \big( n^{-1} \big)  }{ a^{(\mathrm{Type\text{ } III})}\big( \mathrm{exp} \big(1  \big)  \big) -   E^{l,\mathrm{(Type\text{ }III)}}   -\mathrm{o} \big( n^{-1} \big)  }    \big] \text{ } \text{ . } 
\end{align*}

\noindent For $ v^{(\mathrm{Type\text{ } I})},  v^{(\mathrm{Type\text{ } II})}, v^{(\mathrm{Type\text{ } III})} \neq 0$ and $|z| =1$, $z^{-v} \mathcal{W} \big( z \big) $, of each type, can be factorized as,

\begin{align*}
  z^{-v^{(\mathrm{Type\text{ } I})}} \mathcal{W}^{(\mathrm{Type\text{ } I})} \big( z \big)  =  \mathrm{exp} \big[   G^{(\mathrm{Type\text{ } I})}_{-} \big( z \big) G^{(\mathrm{Type\text{ } I})}_{+} \big( z \big)     \big]  \text{ } \text{ , } \\    z^{-v^{(\mathrm{Type\text{ } II})}} \mathcal{W}^{(\mathrm{Type\text{ } II})} \big( z \big)  =  \mathrm{exp} \big[   G^{(\mathrm{Type\text{ } II})}_{-} \big( z \big) G^{(\mathrm{Type\text{ } II})}_{+} \big( z \big)     \big]      \text{ } \text{ , }      \\   z^{-v^{(\mathrm{Type\text{ } III})}} \mathcal{W}^{(\mathrm{Type\text{ } III})} \big( z \big)  =  \mathrm{exp} \big[   G^{(\mathrm{Type\text{ } III})}_{-} \big( z \big) G^{(\mathrm{Type\text{ } III})}_{+} \big( z \big)     \big]   \text{ } \text{ , } 
\end{align*}

\noindent because of the analyticity of $\widetilde{\Psi^{l}}_j$, and of $\widetilde{\Psi^{l}}_i$, with $G_{\pm} \big( z \big) = - \big[ \mathrm{log} \big(   z^{-v} \mathcal{W} \big( z \big)    \big) \big]_{\pm}$. The factorization above is possible when the Toeplitz operator is invertible (see {\color{blue}[3]}). The more general case of invetibility is known to exist from Pollard's Theorem, and the Wiener-Levy Theorem, both of which are referenced in \textit{Section 1} of {\color{blue}[14]}.

From arguments in {\color{blue}[7]}, the factorization mentioned in the previous paragraph holds for $0 < \alpha < | \beta | < 1$, in which case, from the set of equations,

\begin{align*}
   v^{(\mathrm{Type\text{ } I})} = \pm 1 \text{ } \text{ , } \\ v^{(\mathrm{Type\text{ } II})} = \pm  1 \text{ } \text{ , } \\ v^{(\mathrm{Type\text{ } III})} = \pm 1 \text{ } \text{ , } 
\end{align*}

\noindent only $ v^{(\mathrm{Type\text{ } I})} = - 1,   v^{(\mathrm{Type\text{ } II})} = - 1,$ and $v^{(\mathrm{Type\text{ } III})} = -1$ contain nontrivial solutions for $-1 < \beta < 0$. Next, from previous equations of the form,

\begin{align*}
    \mathcal{W} \big( z \big) \widetilde{\Psi^{l}}_j  =    \widetilde{\Psi^{l}}_i     \text{ } \text{ , } 
\end{align*}

\noindent for $|z| = 1$, one has,

\begin{align*}
      \mathrm{exp} \big( - G^{(\mathrm{Type\text{ } I})}_{+} \big( z \big) \big)  \widetilde{\Psi^{l,(\mathrm{Type\text{ } I})}}_j   =  z^{-v^{(\mathrm{Type\text{ } I})}} \mathrm{exp} \big( - G^{(\mathrm{Type\text{ } I})}_{-} \big( z \big) \big)   \widetilde{\Psi^{l,(\mathrm{Type\text{ } I})}}_i \text{ } \text{ , } \\       \mathrm{exp} \big( - G^{(\mathrm{Type\text{ } II})}_{+} \big( z \big) \big)  \widetilde{\Psi^{l,(\mathrm{Type\text{ } II})}}_j   =  z^{-v^{(\mathrm{Type\text{ } II})}} \mathrm{exp} \big( - G^{(\mathrm{Type\text{ } II})}_{-} \big( z \big) \big)   \widetilde{\Psi^{l,(\mathrm{Type\text{ } II})}}_i        \text{ } \text{ , }         \\    \mathrm{exp} \big( - G^{(\mathrm{Type\text{ } III})}_{+} \big( z \big) \big)  \widetilde{\Psi^{l,(\mathrm{Type\text{ } III})}}_j   =  z^{-v^{(\mathrm{Type\text{ } III})}} \mathrm{exp} \big( - G^{(\mathrm{Type\text{ } III})}_{-} \big( z \big) \big)   \widetilde{\Psi^{l,(\mathrm{Type\text{ } III})}}_i \text{ } \text{ . } 
\end{align*}

\noindent The LHS of the equality above is continuous for $|z| \leq 1$ and analytic for $|z| < 1$. Moreover, there exists entire functions $F \big( z \big)$ for which,

\begin{align*}
     \mathrm{exp} \big(  - G^{(\mathrm{Type\text{ } I})}_{-} \big( z \big)   \big) \Psi^{l , (\mathrm{Type \text{ } I})}_j \big( z \big) = F^{(\mathrm{Type \text{ } I})} \big( z \big) \text{ } \text{ , } \\  \mathrm{exp} \big( - G^{(\mathrm{Type\text{ } II})}_{-} \big( z \big)  \big) \Psi^{l , (\mathrm{Type \text{ } II})}_j \big( z \big) = F^{(\mathrm{Type \text{ } II})} \big( z \big)     \text{ } \text{ , }  \\  \mathrm{exp} \big(  - G^{(\mathrm{Type\text{ } III})}_{-} \big( z \big)   \big) \Psi^{l , (\mathrm{Type \text{ } III})}_j \big( z \big) = F^{(\mathrm{Type \text{ } III})} \big( z \big)   \text{ } \text{ , } 
\end{align*}

\noindent over $|z| \leq 1$, for $(\mathrm{Type \text{ } I})$, $(\mathrm{Type \text{ } II})$, and $(\mathrm{Type \text{ } III})$, and,

\begin{align*}
      \mathrm{exp} \big(   G^{(\mathrm{Type\text{ } I})} \big( z \big)    \big)  \Psi^{l , (\mathrm{Type \text{ } I})}_i  = F^{(\mathrm{Type \text{ } I})} \big( z \big)   \text{ } \text{ , }   \\   \mathrm{exp} \big( G^{(\mathrm{Type\text{ } II})} \big( z \big)      \big) \Psi^{l , (\mathrm{Type \text{ } II})}_j = F^{(\mathrm{Type \text{ } II})} \big( z \big) \text{ } \text{ , }   \\    \mathrm{exp} \big(  G^{(\mathrm{Type\text{ } III})} \big( z \big)     \big) \Psi^{l , (\mathrm{Type \text{ } III})}_j = F^{(\mathrm{Type \text{ } III})} \big( z \big)  \text{ } \text{ , } 
\end{align*}

\noindent over $|z| \geq 1$, for $(\mathrm{Type \text{ } I})$, $(\mathrm{Type \text{ } II})$, and $(\mathrm{Type \text{ } III})$. Under the assumption that $v = - 1$,

\begin{align*}
       \Psi^{l , (\mathrm{Type \text{ } I})}_j  =  k^{(\mathrm{Type \text{ } I})}  \mathrm{exp} \big(  G^{(\mathrm{Type\text{ } I})}_{+}   \big( z \big)   \big)   \text{ } \text{ , } \Psi^{l , (\mathrm{Type \text{ } II})}_j = k^{(\mathrm{Type \text{ } II})}  \mathrm{exp} \big( G^{(\mathrm{Type\text{ } II})}_{+}  \big( z \big)     \big) 
 \text{ } \text{ , } \\  \Psi^{l , (\mathrm{Type \text{ } III})}_j   = k^{(\mathrm{Type \text{ } III})}  \mathrm{exp} \big(  G^{(\mathrm{Type\text{ } III})}_{+}     \big( z \big) \big)    \text{ } \text{ , } 
\end{align*}

\noindent for,

\begin{align*}
   F^{(\mathrm{Type \text{ } I})} \big( z \big) =  \overset{|v^{(\mathrm{Type\text{ } I})}| - 1}{\underset{i = -\infty}{\sum}}  k^{\prime}_i z^i  \text{ } \text{ , }  F^{(\mathrm{Type \text{ } II})} \big( z \big) = \overset{|v^{(\mathrm{Type\text{ } II})}| - 1}{\underset{i = -\infty}{\sum}}   k^{\prime\prime}_i z^i\text{ } \text{ , }  F^{(\mathrm{Type \text{ } III})} \big( z \big) = \overset{|v^{(\mathrm{Type\text{ } III})}| - 1}{\underset{i = -\infty}{\sum}} k^{\prime\prime\prime}_i z^i \text{ } \text{ . } 
\end{align*}

\noindent From the value that each $F \big( z \big)$ converges to, which we denote with $k^{(\mathrm{Type \text{ } I})}$, $k^{(\mathrm{Type \text{ } II})}$, and $k^{(\mathrm{Type \text{ } III})}$. From each such $k$, each $G_{+} \big( z \big)$ takes the form,

\begin{align*}
  G^{(\mathrm{Type\text{ } I})}_{+}     \big( z \big) \equiv         - \frac{1}{2 \pi i}  \underset{\textbf{s}}{\oint} \frac{\mathrm{d} z^{\prime} }{z^{\prime} - z} \big(  \mathrm{log} \big( z^{\prime} W^{(\mathrm{Type\text{ } I}} \big( z^{\prime} \big)      \big)_{+}            \text{ } \text{ , } \\ G^{(\mathrm{Type\text{ } II})}_{+}     \big( z \big) \equiv       - \frac{1}{2 \pi i}  \underset{\textbf{s}}{\oint} \frac{\mathrm{d} z^{\prime} }{z^{\prime} - z} \big(  \mathrm{log} \big( z^{\prime} W^{(\mathrm{Type\text{ } II}} \big( z^{\prime} \big)      \big)_{+}                \text{ } \text{ , } \\ G^{(\mathrm{Type\text{ } III})}_{+}     \big( z \big) \equiv       - \frac{1}{2 \pi i}   \underset{\textbf{s}}{\oint} \frac{\mathrm{d} z^{\prime} }{z^{\prime} - z} \big(  \mathrm{log} \big( z^{\prime} W^{(\mathrm{Type\text{ } III}} \big( z^{\prime} \big)      \big)_{+}                   \text{ } \text{ , } 
\end{align*}

\noindent for winding numbers equal to $-1$.

\subsection{Asymptotic evaluation of the contour integral}

\noindent To evaluate each one of the contour integals above which will yield the asymptotic approximation for the eigenvalue of each Fisher-Hartwig symbol, we already know, from {\color{blue}[6]}, that,

\begin{align*}
  \psi^l_j \sim A z_{\mathrm{crit}}^{-(j-1)} + B \big( j + 1 \big)^{-2 \alpha - 1}  \text{ } \text{ , } 
\end{align*}

\noindent for $z_{\mathrm{crit}} \equiv a \big[ \mathrm{exp} \big( - i p^l \big) \big]$. Under the asymptotic approximation, 

\begin{align*}
z_{\mathrm{crit}}^{-n} \approx n^{-2\alpha-1}   \text{ } \text{ , }
\end{align*}

\noindent hence, for $(\mathrm{Type\text{ } I})$, the eigenvector, up to leading first order,

\begin{align*}
 \psi^{l, (\mathrm{Type\text{ } I})}_j - B \big( j + 1 \big)^{-2\alpha -1} \sim A n^{-2 \alpha -1} z_{\mathrm{crit}}^{-1, (\mathrm{Type\text{ } I})} 
 \end{align*}

\noindent is approximately, 
 
 \begin{align*}
 A   n^{-2 \alpha - 1}   \bigg[ 2 - \mathrm{exp}  \big(  \frac{-2 i \pi l}{n}   \big)    - \mathrm{exp}   \big( \frac{-2 i \pi l}{n}   \big)^{-1} \bigg]^{\frac{\alpha}{2}}   \big( - 1 \big)^{\beta} \mathrm{exp}^{\beta}   \big(  \frac{- 2 i \pi l}{n}  \big)     
   +  B \big( j + 1 \big)^{-2\alpha -1}          \text{ } \text{ , } 
\end{align*}

\noindent for $(\mathrm{Type\text{ } II})$, the  eigenvector, up to leading first order,

\begin{align*}
 \psi^{l, (\mathrm{Type\text{ } II})}_j - B \big( j + 1 \big)^{-2\alpha -1} \sim A n^{-2 \alpha -1} z_{\mathrm{crit}}^{-1, (\mathrm{Type\text{ } II})} 
 \end{align*}

\noindent is approximately,

\begin{align*}
   A   n^{-2 \alpha - 1}  \bigg[ 1 - \frac{z_0}{\mathrm{exp} \big( \frac{- 2 i \pi l}{n} \big) } \bigg]^{\delta} \bigg[ 1 - \frac{\mathrm{exp} \big( \frac{- 2 i \pi l}{n} \big)}{z_0}     \bigg]^{\gamma}     b \bigg( \mathrm{exp} \big( \frac{- 2 i \pi l}{n} \big) \bigg)            +  B \big( j + 1 \big)^{-2\alpha -1}          \text{ } \text{ , } 
\end{align*}

\noindent while for $(\mathrm{Type\text{ } III})$, the eigenvector, up to leading first order,

\begin{align*}
 \psi^{l, (\mathrm{Type\text{ } III})}_j - B \big( j + 1 \big)^{-2\alpha -1} \sim A n^{-2 \alpha -1} z_{\mathrm{crit}}^{-1, (\mathrm{Type\text{ } III})} 
 \end{align*}

 \noindent is approximately, 

 \begin{align*}
   A   n^{-2 \alpha - 1}    \overset{M}{\underset{j=1}{\prod}}    \bigg(  \big| \mathrm{exp} \big( - i p^l_j \big) \big|^{2 \alpha_j }    \bigg| \text{ }   \frac{\mathrm{exp} \big( - i p^l \big) }{\mathrm{exp} \big( - i p^l_j \big) }  - 1 \text{ } \bigg|^{2 \alpha_j} \bigg) \text{ }  c \big( \mathrm{exp} \big( - i p^l \big)  \big)      +  B \big( j + 1 \big)^{-2\alpha -1}           \text{ } \text{ , } 
 \end{align*}

\noindent in which one recalls, from \textit{1.3}, for the first type of Fisher-Hartwig symbol,

\begin{align*}
  E^{l,\mathrm{(Type\text{ }I)}} \approx      \bigg[ 2 - \mathrm{exp}  \big(  \frac{-2 i \pi l}{n}   \big)    - \mathrm{exp}   \big( \frac{-2 i \pi l}{n}   \big)^{-1} \bigg]^{\frac{\alpha}{2}}  \big( -1 \big)^{\beta}  \mathrm{exp}^{\beta}   \big(  \frac{- 2 i \pi l}{n}  \big)     
    \text{ } \text{ , } 
\end{align*}

\noindent for the second type of Fisher-Hartwig symbol,

\begin{align*}
 E^{l,\mathrm{(Type\text{ }II)}}  \approx      \bigg[ 1 - \frac{z_0}{\mathrm{exp} \big( \frac{- 2 i \pi l}{n} \big) } \bigg]^{\delta} \bigg[ 1 - \frac{\mathrm{exp} \big( \frac{- 2 i \pi l}{n} \big)}{z_0}     \bigg]^{\gamma}  b \bigg( \mathrm{exp} \big( \frac{- 2 i \pi l}{n} \big) \bigg)           \text{ } \text{ , } 
\end{align*}

\noindent and for the third type of Fisher-Hartwig symbol,

\begin{align*}
  E^{l,\mathrm{(Type\text{ }III)}}  \approx  \overset{M}{\underset{j=1}{\prod}}  \bigg(  \big| \mathrm{exp} \big( - i p^l_j \big) \big|^{2 \alpha_j }    \bigg| \text{ }   \frac{\mathrm{exp} \big( - i p^l \big) }{\mathrm{exp} \big( - i p^l_j \big) }  - 1 \text{ } \bigg|^{2 \alpha_j} \bigg) \text{ }  c \bigg( \mathrm{exp} \big( - i p^l \big)  \bigg)  \text{ } \text{ , } 
\end{align*}

\noindent from which we conclude the argument. \boxed{}

\begin{center}
\begin{tabular}{||c c ||} 
$\text{ Different complex valued functions b}$ \\ 
\hline $  b \big( \mathrm{exp} \big( \frac{- 2 i \pi l}{n} \big) \big)$  & $E^{l,\mathrm{(Type\text{ }II)}}$  \\ 
 \hline
 $\mathrm{exp} \big( - 2 i \pi l\big)  $ &  $\big( 1 - \frac{z_0}{\mathrm{exp} \big( \frac{- 2 i \pi l}{n} \big) } )^{\delta} \big( 1 - \frac{\mathrm{exp} \big( \frac{- 2 i \pi l}{n} \big)}{z_0}     \big)^{\gamma}  \mathrm{exp} \big(  - 2 i \pi l      \big) $        \\
 \hline
 $ \big( 1 - \frac{z_0}{\mathrm{exp} \big( \frac{- 2 i \pi l}{n} \big) } \big)^{-\delta}  $ &   $    \big( 1 - \frac{\mathrm{exp} \big( \frac{- 2 i \pi l}{n} \big)}{z_0}     \big)^{\gamma} $    \\
 \hline
 $ \big( 1 - \frac{\mathrm{exp} \big( \frac{- 2 i \pi l}{n} \big)}{z_0}     \big)^{-\gamma}   $ &  $\big( 1 - \frac{z_0}{\mathrm{exp} \big( \frac{- 2 i \pi l}{n} \big) } )^{\delta}  $        \\ \hline
$ \big( 1 - \frac{z_0}{\mathrm{exp} \big( \frac{- 2 i \pi l}{n} \big) } \big)^{-\delta} \big( 1 - \frac{\mathrm{exp} \big( \frac{- 2 i \pi l}{n} \big)}{z_0}     \big)^{-\gamma} $ & $ b \big( \mathrm{exp} \big( \frac{- 2 i \pi l}{n} \big) \big) $        \\ \hline
$ \bigg[  \big( 1 - \frac{z_0}{\mathrm{exp} \big( \frac{- 2 i \pi l}{n} \big) } \big) \text{ }   \big( 1 - \frac{\mathrm{exp} \big( \frac{- 2 i \pi l}{n} \big)}{z_0}     \big)             \bigg]^{-|\delta + \gamma|}  $ & $ \big(              1 - \frac{z_0}{\mathrm{exp}\big( - 2 i \pi l \big)}       \big)^{- \gamma } \bigg[   \frac{z_0 \mathrm{exp} \big( \frac{-2 i \pi l }{n} \big) - 
 \mathrm{exp} \big( \frac{-2 i \pi l }{n} \big)^2 }{   z_0 \mathrm{exp} \big( \frac{-2 i \pi l }{n} \big) - z^2_0}                  \bigg]^{\delta} $ $  \text{ if } \delta > \gamma$        \\      &  $  \bigg[   \frac{z_0 \mathrm{exp} \big( \frac{-2 i \pi l}{n} \big)   - z_0 }{   z_0 \mathrm{exp} \big( \frac{-2 i \pi l}{n} \big)  -\mathrm{exp}\big(\frac{ - 2 i \pi l}{n} \big)}                  \bigg]^{\gamma} \big(        1 - \frac{\mathrm{exp} \big( \frac{-2 i \pi l}{n} \big) }{z_0}                         \big)^{-\delta} $ $  \text{ if } \delta < \gamma$  \\ 
[1ex] 
 \hline
\end{tabular}
\end{center}

\newpage

\section{References}

\noindent [1] Basor, E.L., Morrison, K.E. The Extended Fisher-Hartwig Conjecture for Symbols with Multiple Jump Discontinuities. \textit{Operator Theory: Advances and Applications} \textbf{71}:16-28 (1994).

\bigskip

\noindent [2] Basor, E.L., Tracy, C.A. The Fisher-Hartwig conjecture and generalizations. \textit{Physics A} \textbf{177}: 167-173 (1991).

\bigskip

\noindent [3] Blackstone, E., Charlier, C., Lenells, J. Toeplitz determinants with a one-cut regular potential and Fisher-Hartwig singularities I. Equilibrium measure supported on the unit circle. \textit{arXiv 2212.06763}.

\bigskip

\noindent [4] Bottcher, A., Silberman, B. Toeplitz Operators and Determinants Generated by Symbols with One Fisher-Hartwig Singularity. \textit{Mathematics Nachrichten} \textbf{127}(1): 95-123 (1986).

\bigskip

\noindent [5] Bottcher, A., Virtanen, J. Norms of Toeplitz Matrices with Fisher-Hartwig Symbolbs. \textit{SIAM Journal on Matrix Analysis and Applications} \textbf{29}(2) (2007). 

\bigskip

\noindent [6] Bottcher, A., Silbermann, B., Widon, H. A Continuous Analogue of the Fisher-Hartwig Formula for Piecewise Continuous Symbols. \textit{Journal of Functional Analysis} \textbf{122}(1): 222-246 (1994).

\bigskip

\noindent [7] Dai, H., Geary, Z., Kadanoff, L.P. Asymptotics of eigenvalues and eigenvectors of Toeplitz matrices. \textit{J. Stat. Phys.}, 05012 (2009).

\bigskip

\noindent [8] Ehrhardt, T., Silbermann, B. Toeplitz Determinants with One Fisher-Hartwig Singularity. \textit{Journal of Functional Analysis} \textbf{148}: 229-256 (1997).

\bigskip

\noindent [9] Kadanoff, L.P. Expansions for Eigenfunction and Eigenvalues of large-n Toeplitz Matrices. \textit{arXiv 0906.0760}.

\bigskip

\noindent [10] Its, A.R., Korepin, V.E. The Fisher-Hartwig Formula and Entanglement Entropy. \textit{Journal of Statistical Physics} \textbf{137}: 1014-1039 (2009).

\bigskip

\noindent [11] Ivanov, D.A., Abanov, A.G. Fisher-Hartwig expansion of Toeplitz determinants and the spectrum of a single-particle reduced density matrix for one-dimensional free fermions. \textit{Journal of Physics A: Mathematical and Theoretical} \textbf{46} (2013).

\bigskip

\noindent [12] Ivanov, D.A., Abanov, A.G., Cheianov, V.V.. Counting free fermions on a line: a Fisher-Hartwig asymptotic expansion for the Toeplitz determinant in the double-scaling limit. \textit{Journal of Physics A: Mathematical and Theoretical} \textbf{46} (2013).

\bigskip

\noindent [13]  Lee,S., Dai, H., and E. Bettelheim, E. Asymptotic eigenvalue distribution of large Toeplitz matrices \textit{arXiv:0708.3124} (2007).

\bigskip

\noindent [14] Movassagh, R., Kadanoff, L.P. Eigenpairs of Toeplitz and disordered Toeplitz matrices with a Fisher-Hartwig symbol. \textit{J. Stat. Phys.} \textbf{167}(3): 959-996 (2017).

\bigskip

\noindent [15] Protopopov, I.V., Gutman, D.B., Mirlin, A.D. Luttinger liquids with multiple Fermi edges: Generalized Fisher-Hartwig conjecture and numerical analysis of Toeplitz determinants. \textit{Lith. J. Physics.} \textbf{52}(2): 165-179 (2012).

\bigskip

\noindent [16] Rambour, P. Orthogonal polynomials with respect to a class of Fisher-Hartwig symbols and inverse of Toeplitz matrices. \textit{Bollettino dell'Unione Matematics Italiana} \textbf{10}: 159-178 (2017).

\bigskip

\noindent [17] Webb, C. A discrete log gas, discrete Toeplitz determinants with Fisher-Hartwig singularities, and Gaussian Multiplicative Chaos. \textit{arXiv 1509.03446}.

\end{document}